\documentclass[universe,article,accept,pdftex,moreauthors]{Definitions/mdpi} 
\firstpage{1} 
\pubvolume{1}
\issuenum{1}
\articlenumber{0}
\pubyear{2026}
\copyrightyear{2026}
\externaleditor{~} 
\datereceived{20 November 2025} 
\daterevised{21 January 2026} 
\dateaccepted{22 January 2026} 
\datepublished{ } 
\hreflink{https://doi.org/} 

\usepackage{graphicx}
\usepackage{caption}
\usepackage{subcaption}
\usepackage{amsmath}
\usepackage{comment}

\Title{Revisiting a Quasar Microlensing Event Towards AGN~J1249{+}3449}

\Author{{Mario Cazzolla} 
 $^{1}$\orcidA{}, Francesco De Paolis $^{1,2,3}$*\orcidB{}, Antonio Franco $^{1,2,3}$\orcidC{} and Achille Nucita $^{1,2,3}$\orcidD{}}

\AuthorNames{Mario Cazzolla, Francesco De Paolis, Antonio Franco, Achille Nucita}

\address{%
$^{1}$ \quad Department of Mathematics and Physics “E. De Giorgi”, University of Salento, Via per Arnesano, CP-I93, {I-73100}~Lecce, Italy; 
 mario.cazzolla@studenti.unisalento.it~(M.C.); {antonio.franco@le.infn.it~(A.F.); achille.nucita@le.infn.it~(A.N.)}\\
$^{2}$ \quad INFN (Istituto Nazionale di Fisica Nucleare), Sezione di Lecce, Via per Arnesano, CP-193, I-73100~Lecce, Italy\\
$^{3}$ \quad INAF (Istituto Nazionale di Astrofisica), Sezione di Lecce, Via per Arnesano, CP-193, I-73100~Lecce, Italy}

\corres{Correspondence: {francesco.depaolis@le.infn.it}}

\abstract{The gravitational wave event GW190521 seems to be the only BH merger event possibly correlated with an electromagnetic counterpart, which appeared about 34 days after the GW event. This work aims to confirm that the electromagnetic bump towards the Active Galactic Nucleus (AGN) J1249+3449 can be explained within the framework of the gravitational microlensing phenomenon. In particular, considering the data of the Zwicky Transient Facility (ZTF), what emerges from a detailed analysis of the observed light curve using three fitting models (Point Source Point Lens, Finite Source Point Lens, Uniform Source Binary Lens) is that the optical bump can be explained as a microlensing event caused by a lens with mass {$\sim\,$0.1 $M_{\odot}$}, lying in the host galaxy of the AGN in question.} 


\begin{document}




\section{Introduction}
\textls[-15]{On 21 May 2019, at~03:02:29 UTC, the~LIGO \citep{Class. Quantum Grav. 1} and VIRGO \citep{Class. Quantum Grav. 2} interferometers revealed a gravitational wave (GW) event, named GW190521, from~a candidate binary black hole (BBH) merger towards AGN J1249+3449, with~a false alarm rate of \mbox{$3.8$ {$\times$} $10^{-9}~\text{Hz}$~\citep{Phys.Rev.Lett. 1}.} 
The~performed analysis allowed constraints on the physical parameters of the \mbox{system \citep{Phys.Rev.Lett. 1, Phys.Rev.Lett. 2},} as~follows:}
\begin{equation}\label{AGN}
    \begin{cases}
        M_1 = 85^{+21}_{-14} \,\,M_{\odot}, \\
        M_2 = 66^{+17}_{-18} \,\,M_{\odot}, \\
        M_{res} = 142^{+28}_{-16} \,\,M_{\odot}, \\
        D_l = 3931 \pm 953\,\,\text{Mpc},
    \end{cases}
\end{equation}
where $M_1$ and $M_2$ represent the masses of the black holes which took part in the merging event, $M_{res}$ is the resulting mass after the black holes coalesce, and~$D_l$ is the luminosity distance of the system. In~\citep{Phys.Rev.Lett. 2} the latter quantity is quantified as $5.3^{+ 2.4}_{- 2.6}\,\text{Gpc}$, but, for~the whole article, the~value estimated in \citep{Phys.Rev.Lett. 1} has been taken into~consideration.

Approximately 34 days after the GW event, an~electromagnetic flare was detected by the Zwicky Transient Facility {(ZTF)}
\endnote{The ZTF is a time-domain optical survey performed by the Palomar 48-inch (1.22 m) Schmidt Telescope. Its camera is characterized by 16 CCDs of 6144 × 6160 pixels each, providing an area of 47 $\text{deg}^2$ for every exposure. The~telescope operates with a limiting magnitude of 20.8 in the $g$ band and~20.6 in the $r$ band (for further details see \citep{Publ. Astron. Soc. Pac. 1}).} \citep{Phys.Rev.D.}. In~\citep{Phys.Rev.Lett. 1}, the~authors associated this flare with a probable electromagnetic counterpart of the merger in the AGN disk, excluding other possible explanations such as intrinsic variability of the AGN, a~supernova, a~tidal disruption event or a microlensing~event.

The latter scenario had been discarded because the expected characteristic timescale for microlensing is in the order of years, which is inconsistent with the several-week flare considered; furthermore, an~event rate estimate was calculated: ``assuming a $M_{\odot}$ lens in the source galaxy, the~lens is required to orbit at $\sim$1 kpc with a transverse velocity of about 200~$\text{km}/ \text{s}$ in order to match the timescale ($\sim$$2$ {$\times$} $10^6\,\text{s}$) and magnification ($\sim$1.4) of this event; assuming a population of $O (10^{10})$ stars in appropriate orbits, geometric considerations produce a rate of $O(10^{-5})$ events $\text{yr}^{-1}$ $\text{AGN}^{-1}$\,'' \citep{Phys.Rev.Lett. 1}. 

Just after the discovery of the electromagnetic flare, a~discussion of its nature began in the astronomical community, since no electromagnetic counterpart is expected for black hole mergers.
Indeed, in~\citep{Mon. Not. R. Astron. Soc. 1} it is shown that the detected optical light curve is well fitted by a microlensing event due to a lens with mass about $ 0.1\, M_{\odot}$ lying in the AGN disk. The~analysis allowed for constraining the system parameters, as~follows:
\begin{equation}
    \begin{cases}
        t_0 = 469\, \pm \,1 \,\text{days}, \\
        t_E = 17.6 \,\pm\,1.2\,\text{days}, \\
        u_0 = 0.92\,\pm\, 0.03, \\
        m_b = 19.082 \,\pm\,0.003.
    \end{cases}
\end{equation}
{Here,} 
$t_0$ is the time of closest approach between the lens and the source, and~it is expressed considering $MJD = 58,202.293102$ (the start of data acquisition, expressed in Modified Julian Date) as $t = 0$. $u_0$ is the impact parameter in units of the Einstein radius, $m_b$ is the magnitude of the source baseline (the intrinsic flux of the source, in~the absence of the microlensing event), and~$t_E$ is the Einstein crossing time:
\begin{equation}\label{formula_EA}
t_E = \frac{r_E}{v_{\perp}} = \frac{\theta_E D_L}{v_{\perp}},
\end{equation}
{where} 
$v_{\perp}$ is the relative transverse velocity between the source and lens due to the proper motions of Earth, lens and source, $r_E$ and $\theta_E$ are the physical and angular Einstein ring radius, respectively, and~$D_L$ is the distance to the~lens. 

As mentioned in~\citep{Mon. Not. R. Astron. Soc. 1}, detecting such transient events is very challenging, so it is necessary to acquire a large amount of data to understand these events better. Therefore, the~aim of this article is to study the AGN light curve coming from the AGN with more recent data (in order to show that, for~the entire light curve, there are no periodic repetitions of the bump, which can be due to the activity of the quasar), since the above-mentioned articles~\citep{Phys.Rev.Lett. 1, Phys.Rev.Lett. 2, Mon. Not. R. Astron. Soc. 1} considered data until 2019. 
For this reason, in~the present paper, a~new set of data (the ``data release 23'' of the ZTF), with~$g$, $r$, and~$i$ SDSS magnitudes, which covers a time interval going from March 2018 to October 2024, has been taken into~consideration. 

\section{Event~Modelization}
Throughout the paper, the~ZTF light curve is fitted by using three microlensing models, each with a different number of free~parameters:
\begin{itemize}
    \item Point Source Point Lens (PSPL, Section~\ref{PSPL_Model}): fitted through the Levenberg--Marquardt (LM) algorithm, the~Differential Evolution (DE) algorithm and the Markov Chain Monte Carlo (MCMC) method.
    \item Finite Source Point Lens (FSPL, Section~\ref{FSPL_Model}): fitted through the Differential Evolution and the Markov Chain Monte Carlo method. The~LM method is not applicable for this and for the next model because the function ``LM\textunderscore fit'' of pyLIMA\endnote{In this work, we have used the version number 1.9.7 of pyLIMA.} is not compatible with them (for further details see \citep{AJ 1}).
    \item Uniform Source Binary Lens (USBL, Section~\ref{USBL_Model}): fitted using the same algorithms as the FSPL model. 
\end{itemize}
In the following, only the results obtained with the MCMC method will be illustrated, as~this algorithm yields more precise outcomes. Indeed, the~starting parameters are taken from the DE results, and~the latter are taken from the LM method. The~process of considering the previous results as the starting point for the next model has been followed since it allows for faster convergence of the fit. Moreover, the~MCMC algorithm returns more realistic and stable estimations of the uncertainties than the other methods; however, this accuracy makes the MC algorithm about 2000 times slower with respect to the LM method, since it requires the computation of thousands of models \citep{AJ 1}.

\subsection{PSPL~Model}\label{PSPL_Model}
\begin{figure}[H]
    \includegraphics[scale = 0.5]{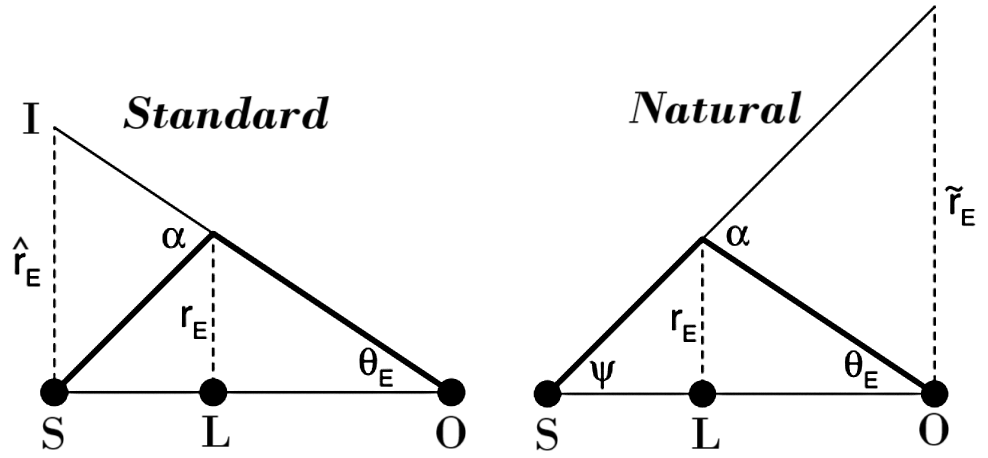}
    \caption{(\textbf{Left Panel}): standard microlensing geometry; the bold curve shows the path of the light from the source (S) to the observer (O) being deflected by the lens (L) of mass $M_L$. The~image (I) is displaced from the source by the angular Einstein radius $\theta_E$, which, projected onto the source plane, corresponds to a physical distance $\hat{r}_E$. (\textbf{Right Panel}):  natural microlensing geometry; mostly the same as the left panel except that the Einstein radius is now projected onto the observer plane as $\tilde{r}_E$ rather than onto the source plane as $\hat{r}_E$. Figure adapted from \citep{ApJ 1}.
}
    \label{microlensing_geometry}
\end{figure}

Since the non-negligible redshift of the source generates differences between the comoving distance, the~luminosity distance and the angular distance, in~this paragraph, the~description made by Gould \citep{ApJ 1} will be followed using the angular distances (see \citep{Mon. Not. R. Astron. Soc. 2}). In~microlensing events, the~lens object is, in~general, a~star nearly crossing the line of sight between the observer and a background source.
Inverting the standard microlensing geometry, i.e.,~projecting the Einstein ring onto the observer plane rather than the source plane, an~evident relation arises between the observable $\theta_E$ and $\tilde{r}_E$, the~projected Einstein radius, and~the physical quantities $M_L$ and $\pi_{rel}$, the~lens mass and the lens--source relative parallax, respectively (see Figure~\ref{microlensing_geometry}).

\vspace{-6pt}

In the new geometry, using the small-angle approximation ($\tan \theta \approx \sin \theta \approx \theta$), one can note that
\begin{equation}
\frac{\alpha}{\tilde{r}_E} = \frac{\theta_E}{r_E} \Longleftrightarrow \theta_E \,\tilde{r}_E = \alpha\,r_E = \frac{4GM_L}{c^2}.    
\end{equation}
{Next,}
\begin{equation}
\theta_E = \alpha - \psi = \frac{\tilde{r}_E}{D_L} - \frac{\tilde{r}_E}{D_S} = \frac{\tilde{r}_E}{D_{rel}},
\end{equation}
in which $D_S$ is the distance to the source and $D^{-1}_{rel} \equiv D^{-1}_L - D^{-1}_S$. The~equation for the angular Einstein radius can be rewritten as

\begin{equation}\label{parallax}
    \pi_E\,\theta_E = \pi_{rel},\,\,\,\,\pi_E \equiv \frac{\text{AU}}{\tilde{r}_E},
\end{equation}
where $\pi_{rel} \equiv \frac{\text{AU}}{D_{rel}}$ is the lens--source relative parallax, and~$1\, \text{AU} \equiv 1.496$ {$\times$} $10^8\,\text{km}$ indicates the astronomical~unit.

{From} 
these equations, one gets
\begin{equation}\label{parameters}
\tilde{r}_E = \sqrt{\frac{4GM_LD_{rel}}{c^2}},\,\,\,\,\,\,\,\pi_E = \sqrt{\frac{\pi_{rel}}{\kappa M_L}},\,\,\,\,\,\,\,\theta_E = \sqrt{\frac{4GM_L}{c^2 D_{rel}}} = \sqrt{\kappa M_L \pi_{rel}},    
\end{equation}
where $\kappa \equiv \frac{4G}{AU c^2} \simeq 8.144\, \frac{mas}{M_{\odot}}$\,. The~standard parametrization in terms of the microlensing parallax vector components takes into account the East and North components, i.e.,~$\pi_{EE}$ and $\pi_{EN}$ respectively, whose composition gives the $\pi_E$ vector absolute value:
\begin{equation}
\pi_E = \sqrt{\pi^2_{EE} + \pi^2_{EN}}.    
\end{equation}

{Combining} the first expression of Equation~(\ref{parallax}) and the last of Equation~(\ref{parameters}), it is possible to obtain an equation for the lens mass:
\begin{equation}
M_L = \frac{\theta_E}{\kappa \pi_E}.
\end{equation} 

{Each} image is magnified and the source flux increases by the total magnification factor $\mu(t)$, which, for~the PSPL model, is given by Paczy\'nski \citep{ApJ 2}:
\begin{equation}\label{magnification}
\mu_{PSPL} = \frac{u(t)^2 + 2}{u(t)\sqrt{u(t)^2 +4}},\,\,\,\,\,u(t) = \sqrt{u^2_0 + \frac{(t-t_0)^2}{t^2_E}},
\end{equation}
where $u(t)$ is the source--lens impact parameter and $u_0 = u(t_0)$ is the minimum impact parameter at the time $t_0$.

\subsection{FSPL~Model}\label{FSPL_Model}
In the PSPL model, it is assumed that the source is a point-like object. However, this hypothesis breaks down when the source-lens separation becomes small enough to be comparable to (or not negligible with respect to) the normalized angular source radius $\rho = \frac{\theta_S}{\theta_E}$, where $\theta_S$ is the angular radius of the source. ``This indicates that finite source effects often appear in the case of highly magnified events'' \citep{AJ 1}. 
In the scenario of an extended source, assuming that it emits homogeneously up to the angular radius $\rho$, the~total amplification is calculated by integrating the amplification $\mu_{tot}$ over the source area S \citep{ApJ 3}:
\begin{equation}
\mu^*(u, \rho) = \frac{\int_S \mu(t)\,\text{d}S}{\int_S \text{d}S} = \frac{1}{\pi \rho^2}\int_S \mu(t)\,\text{d}S,
\end{equation}
in which $\mu(t)$ is given by the PSPL model in Equation~(\ref{magnification}).

{Since} $\rho$ is defined as the source size in units of the Einstein angle, estimating the $\rho$ value makes it easier to estimate the Einstein angle through the relation:
\begin{equation}\label{source_radius}
\theta_E = \frac{R_S}{\rho D_S},  
\end{equation}
where $R_S$ is the source~radius.

\subsection{USBL~Model}\label{USBL_Model}
\begin{figure}[H]
\includegraphics[scale=0.287]{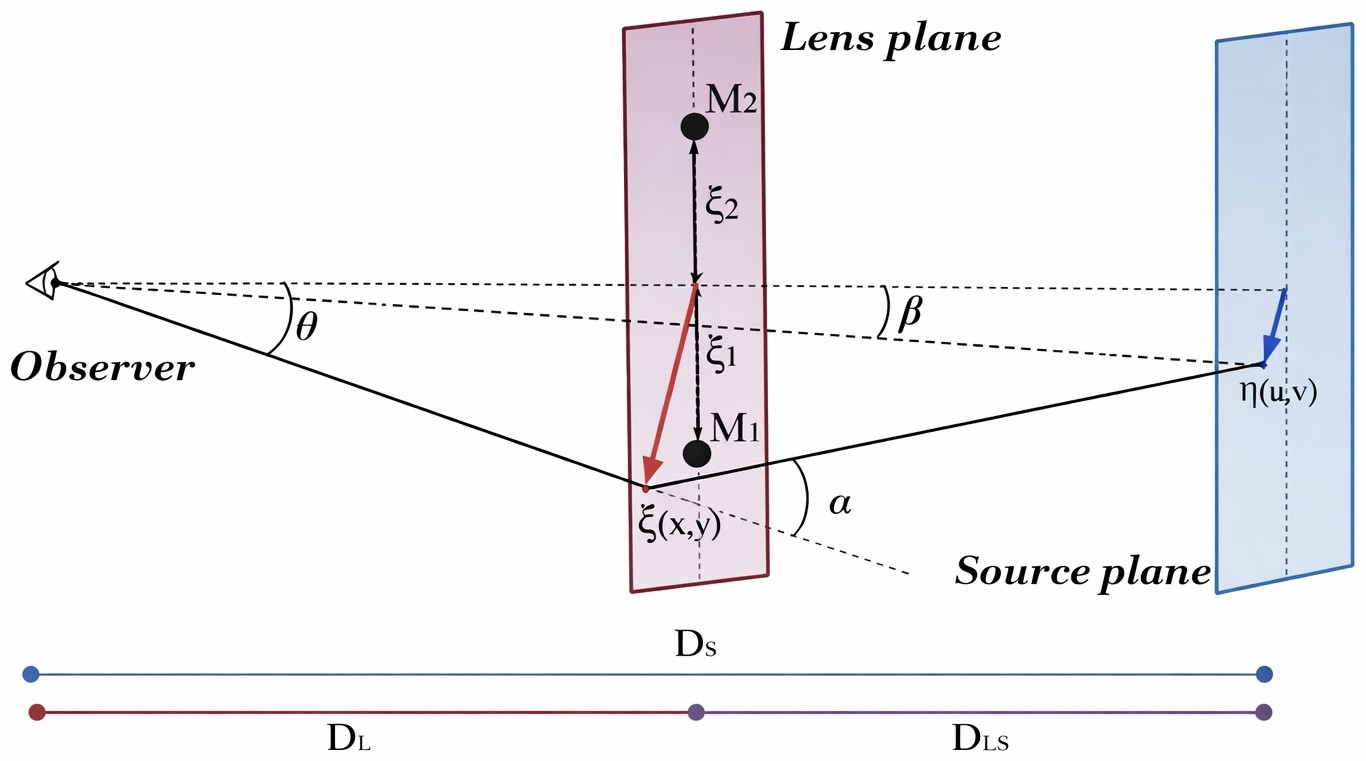}
\caption{{The} 
geometry of a binary lens system. The~lens plane is defined by the coordinate system $(x,y)$, with~the $x$-axis passing through the projection of the two masses onto the plane itself, with~the central point between the lens objects. \textls[-15]{The~source plane is, instead, expressed by the coordinate system $(u,v)$, presenting its origin at the intersection point between the plane and the optical axis. $\xi_1$ and $\xi_2$ are the vector positions of the projection points of the masses ($M_1$ and $M_2$) in the lens plane geometry; $\alpha$ represents the deflection angle, $\beta$ indicates the angular position of the point $\eta (u, v)$ in the source plane in which the light ray originates, and~$\theta$ is the angular position of the intersection point $\xi (x, y)$ between the light ray and the lens plane. The~distances are expressed as follows: $D_L$ is the distance between the observer and the lenses, $D_{LS}$ stands for the distance between the binary lens and the source, and~$D_S$ indicates the distance between the observer and the source.
Figure adapted from \citep{Geosci. 1}.}
}
\label{fig:Binary_lenses}
\end{figure}

A binary lens consists of a two-mass system, composed of two objects with masses $M_1$ and $M_2$. The~lens plane coordinate system $(x,y)$ is arranged so that the $x$-axis passes through the projection of the masses onto the lens plane, with~the origin at the central point between the two lens masses (Figure \ref{fig:Binary_lenses}). Each member of the binary system is considered to be at a distance $D_L$ from the observer, with their separation being relatively~small.
The source plane is described by the coordinate system $(u,v)$, with~its origin at the point where the optical axis intersects the plane. Each mass deflects a light ray from the source by the angle $\alpha = \frac{4GM_L}{c^2b}$ \citep{Sci. 1, Annu. Rev. of Astron. and Astrophys. 1}, where $G$ is the Gravitational Constant and $b$ is the impact parameter between the photon and the lens mass. Consequently, the~total deflection is then the vector sum of the deflections by each of the individual~masses. 

{For} this study, the~light curve of a microlensing event due to a binary lens system can be modeled using seven parameters: the mass ratio $q$, the~binary separation $s$, the~time $t_0$ of the projected closest separation between the source and the binary center of mass, the~Einstein time $t_E$, the~impact parameter $u_0$ with respect to the center of mass, the~source finite size $R_S$, and the~source trajectory angle $\gamma$ with respect to the binary lens axis.
The light curve for a generic binary lens system derives from the solution of the lens equation, which, in~complex notation, can be written as follows \citep{Astron. and Astrophys. 1, ApJ 4, Mon. Not. R. Astron. Soc. 3}:
\begin{equation}
    \eta = \xi + \frac{M_1}{\xi_1 + \overline{\xi}} + \frac{M_2}{\xi_2 + \overline{\xi}}
\end{equation}
where $\eta = u + iv$ and $\xi = x + iy$ are the positions of the source and image, respectively, and~$\xi_1$ and $\xi_2$ are the positions of the two lenses.
The shape of the light curve depends on the binary separation or, more
precisely, on~the projected separation into the sky distance between the two components at the time of interest, expressed in Einstein radius units, $s = \frac{a_{\perp}}{D_L\theta_E}$, where $a_{\perp}$ represents the projected physical separation between the lens masses expressed in~AU. 

\section{Light Curve~Analysis}\label{Light_curve_analysis}
\begin{figure}[H]
    \includegraphics[scale = 0.326]{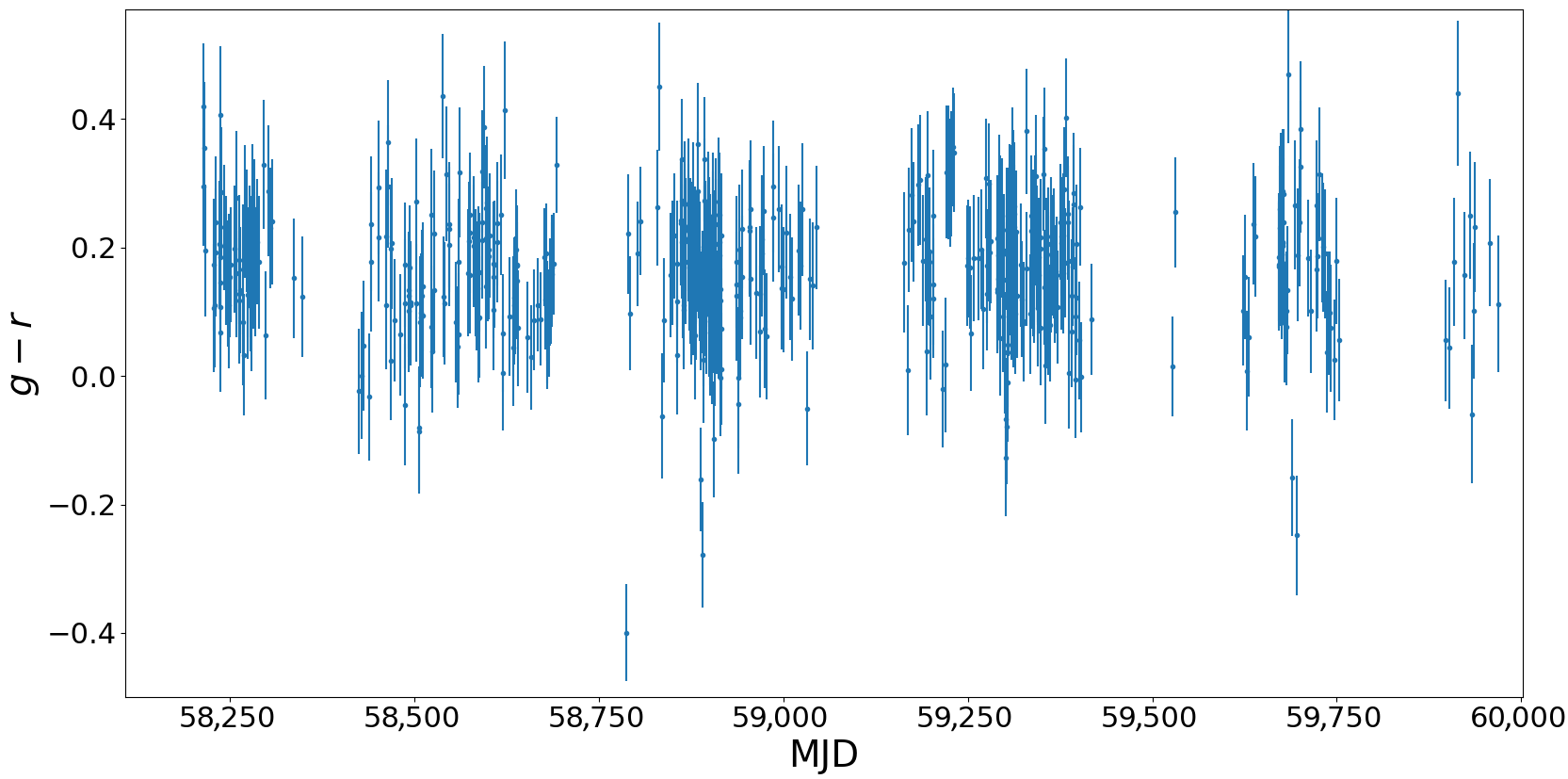}
    \caption{{The} 
    $g-r$ color for the AGN J1249+3449 light curve is shown on the vertical axis, while the horizontal axis represents the time expressed in MJD. For~this figure, the~$r$ band magnitudes have been interpolated at the time of the $g$ band measurements. The~range of time ends at approximately 60,000 since it includes the whole microlensing event and does not consider the last group of points, where the concentration of $r$ band data is very low. The~magnitude remains constant for the entire~graph.}
    \label{interpolation}
\end{figure}

A microlensing event fulfills the following~properties:
\begin{itemize}
    \item The observed light curve is achromatic since the gravitational bending of light is independent of the frequency of the radiation (even though the size and position of the source could shift a bit between the different energies, causing effects that depend on the wavelength).
    \item For a single lens event, the~observed bump should be symmetric with respect to its maximum, and~the overall shape depends on four parameters: the Einstein crossing times $t_E$, the~impact parameter $u_0$, the~time of closest approach $t_0$ and the magnitude of the source baseline $m_b$.
    \item Outside the microlensing event, the~source is characterized by a constant flux, and~the light curve does not present any periodic feature, unless~in the case of a binary source or a binary lens with orbital time much less than the microlensing event duration (for details see, e.g.,~\citep{IJMPD 1, Mon. Not. R. Astron. Soc. 4}).
\end{itemize}
\textls[-20]{In Figure~\ref{interpolation}, the~$g-r$ color for the AGN J1249+3449 light curve is shown (the $i$ band has not been considered due to the small number of data points). Through the interpolation, it is possible to associate a common baseline for the $r$ and $g$ band data, since they have, a~priori, different baselines. As~one can see, the~overall behavior of the $g-r$ light curve is almost achromatic (constant with time), giving further support to the microlensing nature of the considered~event.}
\unskip

\section{Results}
As mentioned above, the~light curve was fitted with three different models (six if the parallax effect was included in each of them) through the use of the open-source software {pyLIMA} 
\citep{AJ 1}: the first couple considered was that formed by the PSPL model and the PSPL + parallax model; subsequently, more realistic and complex models, i.e.,~the FSPL model and the FSPL + parallax model, were taken into consideration; the last situation treated was that of an event caused by a binary lens, through the inclusion of the USBL and USBL + parallax~models.

{Tables}~\ref{table:1}--\ref{table:3} show the models' parameters and their corresponding values, obtained using the MCMC fitting algorithm.

\begin{table}[H]

\caption{{Parameters} 
of the adopted models with their units and a brief~description.}
\label{table:1}
\begin{tabularx}{\textwidth}{cc C } 
 \toprule
 \textbf{Parameter} & \textbf{Unit} & \textbf{Description}  \\ 
 \midrule
 $t_0$ & days & Time of the minimum impact parameter  \\  \midrule
 $u_0$ & $\theta_E$ & Minimum impact parameter defined on the center of mass of the lens system (positive if the source passes on the left of the~source)\\ \midrule

 $t_E$ & days & Angular Einstein ring crossing time  \\ \midrule
 $\rho$ & $\theta_E$ & Normalized angular source radius \\ \midrule
 $s$ & $\theta_E$ &  Normalized angular separation between the binary lens component \\  \midrule
 $q$ & & Binary mass ratio\\ \midrule
 $\gamma$ & rad & Angle between the source trajectory and the binary lens axis (anti-clockwise)\\ \midrule
 $\pi_{EN}$ & $\text{AU}/r_E$ & North component of the microlensing parallax\\ \midrule
 $\pi_{EE}$ & $\text{AU}/r_E$ & East component of the microlensing parallax\\
 \bottomrule
\end{tabularx}
\end{table}
\unskip

\begin{table}[H]
 
    \caption{Obtained values for the parameters in the case of non-parallax models. The~given errors have been estimated by the MCMC fitting~algorithm. The * symbol is used when a parameter is not included in the corresponding model.}
    \label{table:2}
    \begin{tabularx}{\textwidth}{C C C C}
    \toprule
    & \textbf{PSPL} & \textbf{FSPL} & \textbf{USBL} \\
    \midrule
$t_0$ &  58,670.64$^{+0.93}_{-0.79}$ & 58,670.91$^{+1.00}_{-0.95}$& 59,084.1$^{+4.1}_{-9.7}$\\

$u_0$ &  $0.17^{+0.58}_{-0.94}$ & 
$0.44^{+0.36}_{-0.89}$ & $0.09\pm 0.01$\\

$t_E$ & $23.9^{+10.0}_{-4.3}$ & $25.7^{+11.0}_{-5.5}$ &  $380.2^{+8.9}_{-17.0}$\\

$\rho$ & {*} 
& $0.03 \pm 0.02$ & $0.26^{+1.80}_{-0.24}$ {$\times$} $10^{-3}$\\

$s$ & * & * & $2.51 \pm 0.06$\\

$q$ & * & * & $0.98^{+0.02}_{-0.04}$\\

$\gamma$ & * & * & $3.09 \pm 0.01$ \\

$\chi^2$ & $2967.13$ & $2967.14$ & $2418.94$ \\

$\chi^2 / dof$ & $1.86$ & $1.86$ & $1.53$\\ \bottomrule
    \end{tabularx}
\end{table}
\unskip

\begin{table}[H]
 
    \caption{Obtained values for the parameters in the case of parallax models. The~given errors have been estimated by the MCMC fitting~algorithm. The * symbol is used when a parameter is not included in the corresponding model.}
    \label{table:3}
    \begin{tabularx}{\textwidth}{C C C C}
    \toprule
    & \textbf{PSPL + Parallax} & \textbf{FSPL + Parallax} & \textbf{USBL + Parallax} \\
    \midrule
$t_0$ &  58,630$^{+72}_{-41}$ & 58,647$^{+65}_{-93}$& 58,633$^{+42}_{-30}$ \\

$u_0$ &  $-0.05^{+0.72}_{-0.69}$ & 
$0.15^{+0.66}_{-0.84}$ & $0.06^{+0.55}_{-0.74}$ \\

$t_E$ & $28^{+14}_{-8}$ & $29^{+18}_{-10}$ &  $21.9^{+3.8}_{-3.7}$ \\

$\rho$ & {*} 
& $0.02 \pm 0.02$ & $0.5^{+7.3}_{-0.4}$ {$\times$} $10^{-3}$ \\

$s$ & * & * & $0.65^{+1.05}_{-0.10}$ \\

$q$ & * & * & $0.29^{+0.18}_{-0.15}$ \\

$\gamma$ & * & * & $5.41^{+0.13}_{-0.14}$ \\

$\pi_{EN}$ & $0.08^{+0.11}_{-0.14}$& $-0.09^{+0.16}_{-0.10}$& $0.07 \pm 0.09$ \\

$\pi_{EE}$ & $-0.23^{+0.41}_{-0.16}$& $-0.18^{+0.40}_{-0.26}$& $-0.36^{+0.37}_{-0.11}$\\

$\chi^2$ & $2963.32$ & $2958.23$& $6432.78$ \\

$\chi^2/dof$ & $1.85$ & $1.85$ & $4.02$\\ \bottomrule
    \end{tabularx}
\end{table}

As can be seen from Table~\ref{table:3}, for~all the parallax models, the~two components $\pi_{EN}$ and $\pi_{EE}$ are compatible with 0. 
This result is due to the fact that the event occurs over tens of days, whereas a non-negligible parallax contribution is obtained in a few months. {Furthermore, the~USBL model presents the greatest $\chi^2$ value since the data do not properly constrain it, resulting in an unrealistic model}. For~this reason, the~analysis was conducted considering only non-parallax~models.

{In} the case of the latter models, it was possible to calculate the Einstein angle $\theta_E$ using Equation~(\ref{formula_EA}), after~accounting for the lens masses and separation (for the USBL model). 
Since the geometry of the event suggests the presence of the lens within the lens host galaxy~\citep{Mon. Not. R. Astron. Soc. 1}, we chose a range of the lens--source distance $D_{LS}$ between $5 \,\text{kpc}$ and $100 \,\text{kpc}$. Indeed, it can be expected that the lens object is not too close to the center of the galaxy, occupied by a Supermassive Black Hole (SMBH) of $10^{8-9} \, M_{\odot}$, nor too far from it, since the distribution of celestial bodies (which eventually could generate a microlensing event) in the galaxy decreases with the distance from its~center. 

\begin{figure}[H]
    \includegraphics[scale = 0.625]{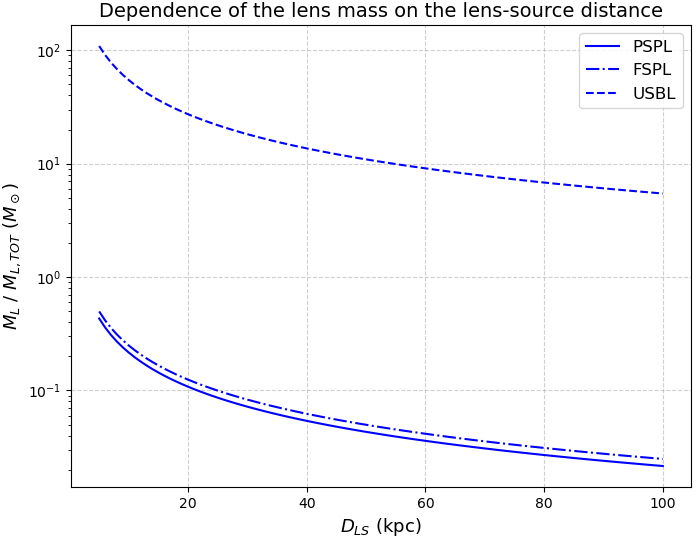}
    \caption{Dependence of the lens mass $M_L$ in $M_{\odot}$ (for PSPL and FSPL models) and of the total lens mass system $M_{L,\,TOT}$ in $M_{\odot}$ (for the USBL model) with respect to the lens--source distance $D_{LS}$ (in kpc). The \emph{y}-axis, rescaled logarithmically, represents the lens mass, while the distance is graphed in the linear \emph{x}-axis. These two parameters are linked by the equation for the calculation of the Einstein time of the event $t_E$ (Equation \eqref{formula_EA}), which can be rewritten as follows (considering $D_L \simeq D_S$): $M_L = \frac{(t_E\, c \,v_{\perp})^2}{4 G D_{LS}}$. The~PSPL model is characterized by a continuous line, the~FSPL model by a dash-dotted line, and~the USBL model by a dashed~line.}
    \label{fig:Lens_Mass_Dist.}
\end{figure}

{Then}, assuming a transverse velocity $v_{\perp} = 300 \, \text{km/s}$, the~lens mass is plotted as a function of $D_{LS}$ (see Figure~\ref{fig:Lens_Mass_Dist.}) for the three considered~models.

In order to present a realistic system and geometry of the event, a~reasonable value of $D_{LS}$ was chosen for the analysis; this value is related to the probability that the microlensing event can occur, thanks to the presence of celestial objects in the galaxy.
The distance chosen was 20 kpc, and~with it, the~calculations of mass and separation (in the USBL model) were carried out. 
The values obtained are presented in Table~\ref{table:4}.

As can be seen from Table~\ref{table:2}, the~introduction of the parameter $\rho$ for the FSPL model does not involve a better estimation of the $\chi^2$; furthermore, its value is not significant enough to make this model more convincing than the PSPL model.
\begin{table}[H]

    \caption{Best fit values of the lens masses and their relative separation $a_{\perp}$  obtained by the PSPL, FSPL and USBL models, considering $D_{LS}$ = 20,000  $\text{pc}$ and $v_{\perp} = 300 \,\text{km/s}$. The * symbol is used when a parameter is not included in the corresponding model. The~errors are calculated using the standard formula for the propagation of uncertainty of a function $f(x, y, \dots)$: $\sigma_f = \sqrt{\Big|\frac{\partial f}{\partial x}\Big|^2\sigma^2_x + \Big|\frac{\partial f}{\partial y}\Big|^2\sigma^2_y + \,\dots}$.}
    \label{table:4}
    \begin{tabularx}{\textwidth}{C C C C}
    \toprule
    & \textbf{PSPL} & \textbf{FSPL} & \textbf{USBL} \\
    \midrule
   $M_{L1}\,(M_{\odot})$ & {$0.10$} & {$0.12$}& {$13.2$}\\
   $M_{L2}\,(M_{\odot})$ & {*} 
   & * & {$13.4$}\\
   $a_{\perp}\,(\text{AU})$ & * & * & $162 \pm 57$\\\bottomrule
    \end{tabularx}
\end{table}

\begin{figure}[H]
    \includegraphics[scale = 0.275]{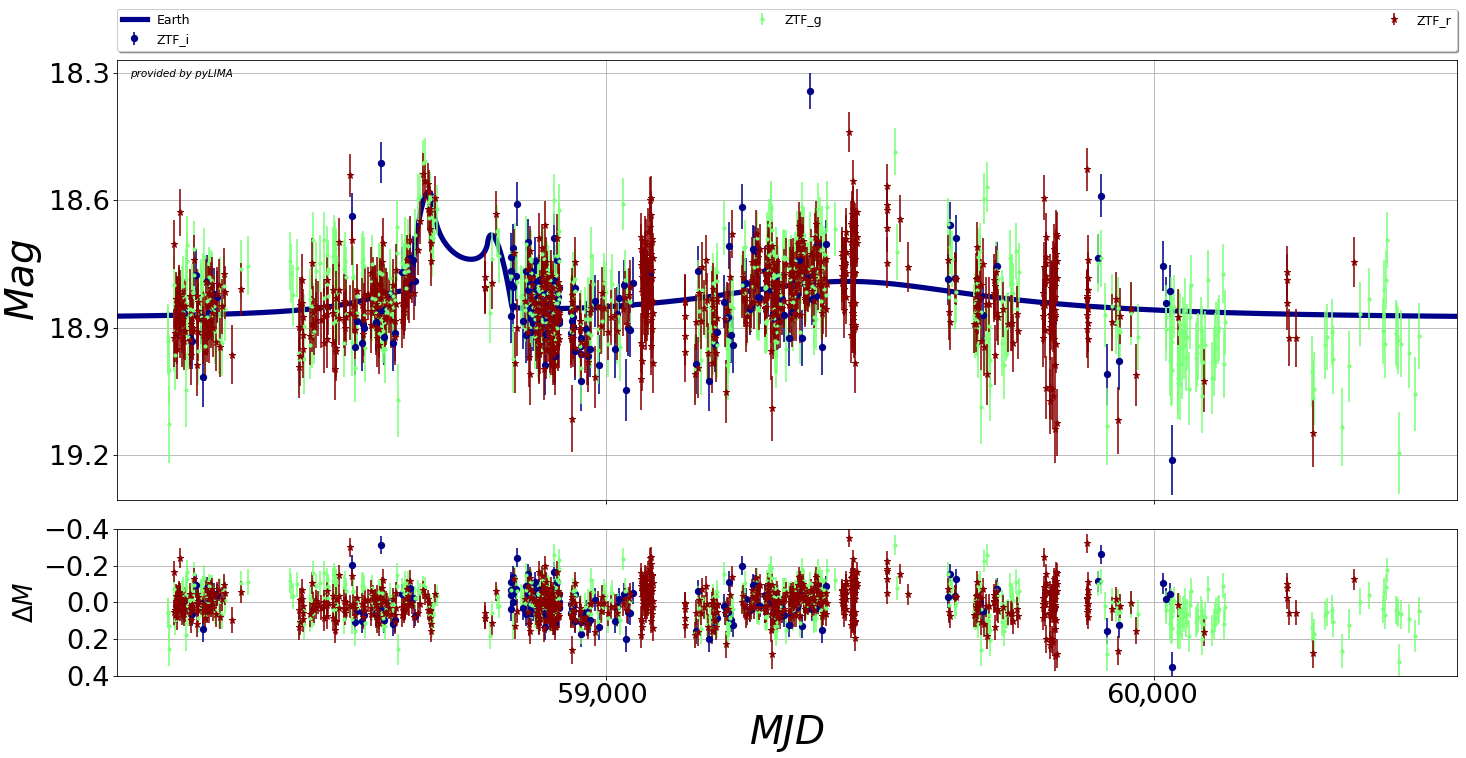}\\\vspace{-8pt}

    \includegraphics[scale = 0.275]{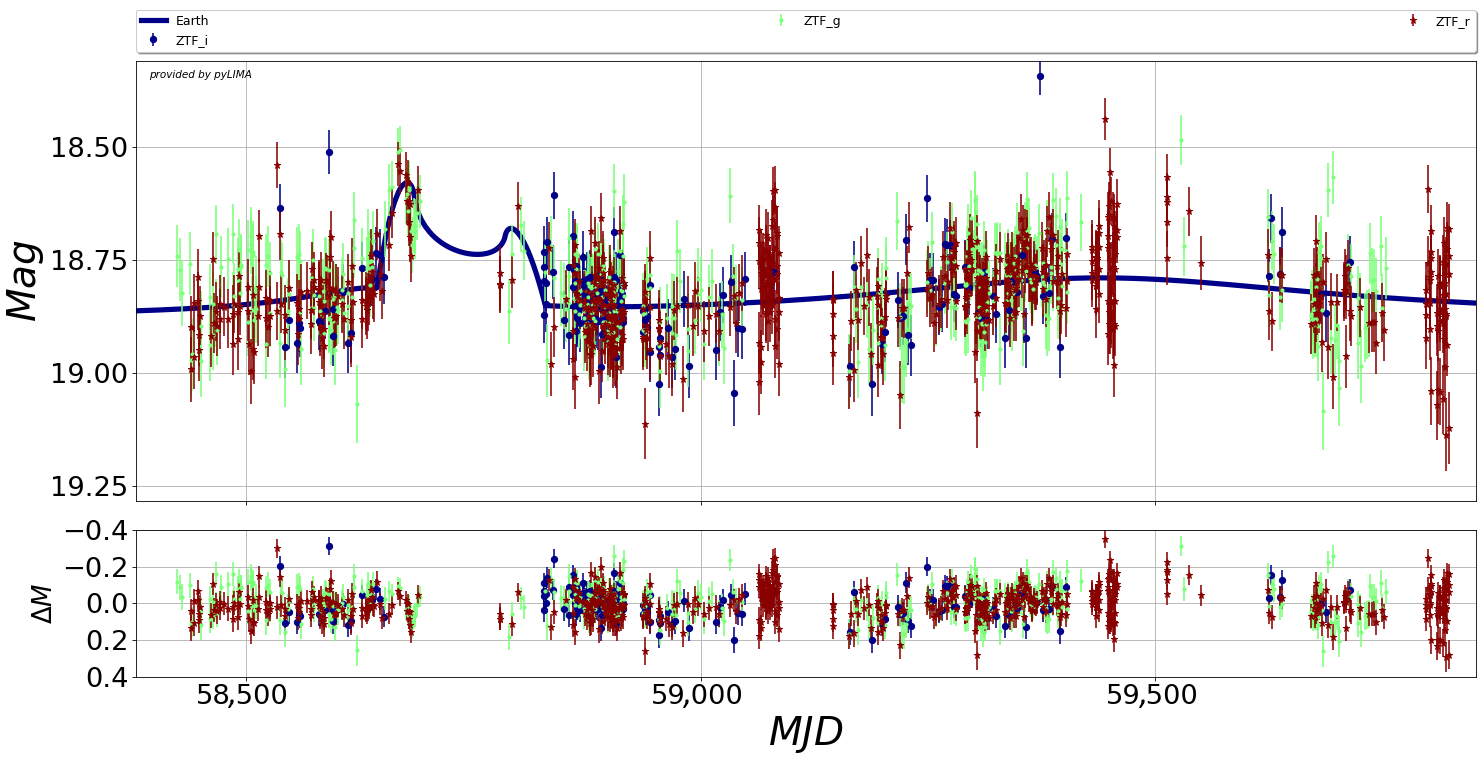}
    \caption{(\textbf{Upper panel}): {Light} 
    curve of the AGN J1249+3449 fitted with the USBL model and the model residuals with respect to the best fit, with~the vertical axis showing the magnitude of the source. (\textbf{Lower panel}): Close-up of the light curve and the model residuals, with~the magnitude expressed on the vertical axis. For~both panels, the~horizontal axis represents the time expressed in MJD.
    The green points indicate the data in the $g$ band, the~red ones the data in the $r$ band, and~the blue points represent the data in the $i$ band. All the magnitudes are intrinsically corrected, with~respect to a common baseline, by~pyLIMA.}
    \label{USBL_fit}
\end{figure}

\begin{figure}[H]
\includegraphics[scale = 0.275]{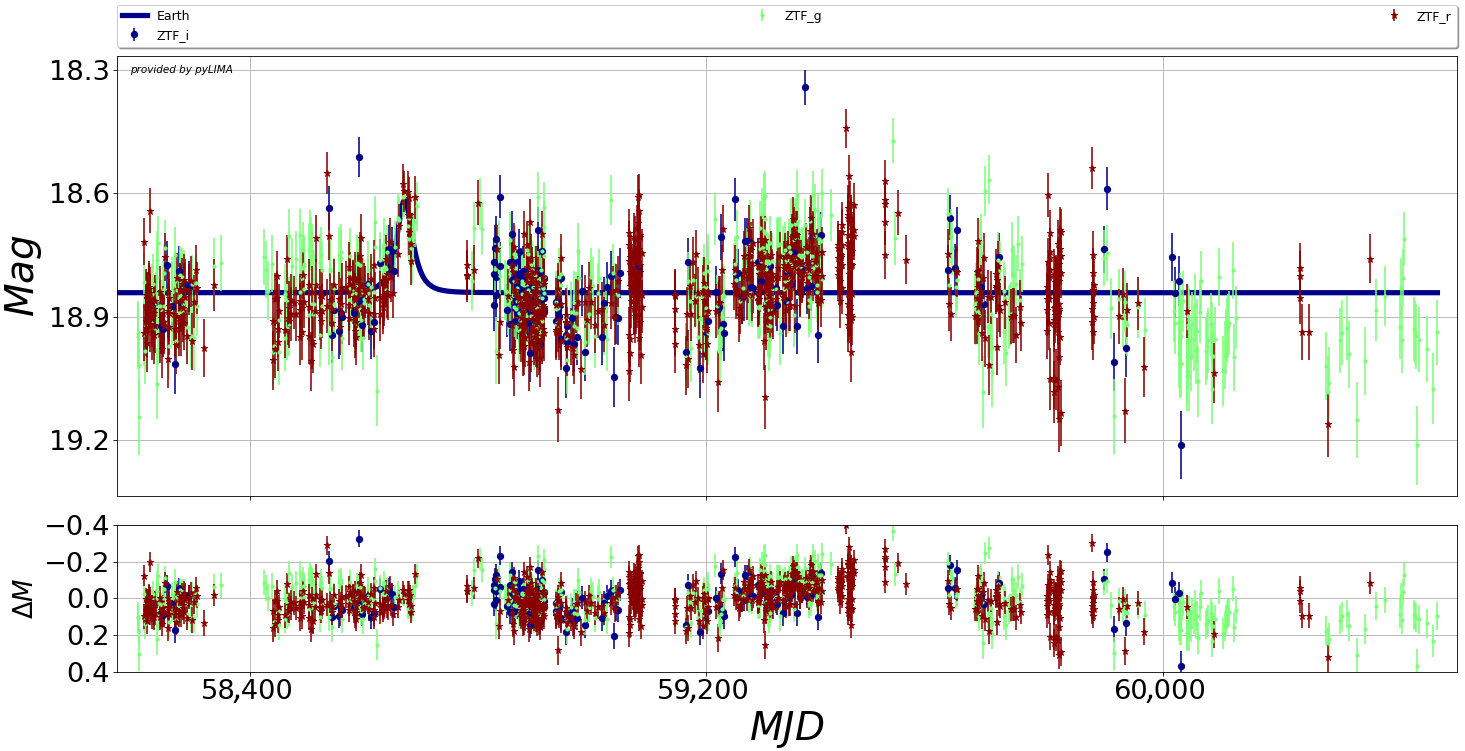}
\\\vspace{-8pt}

\includegraphics[scale = 0.275]{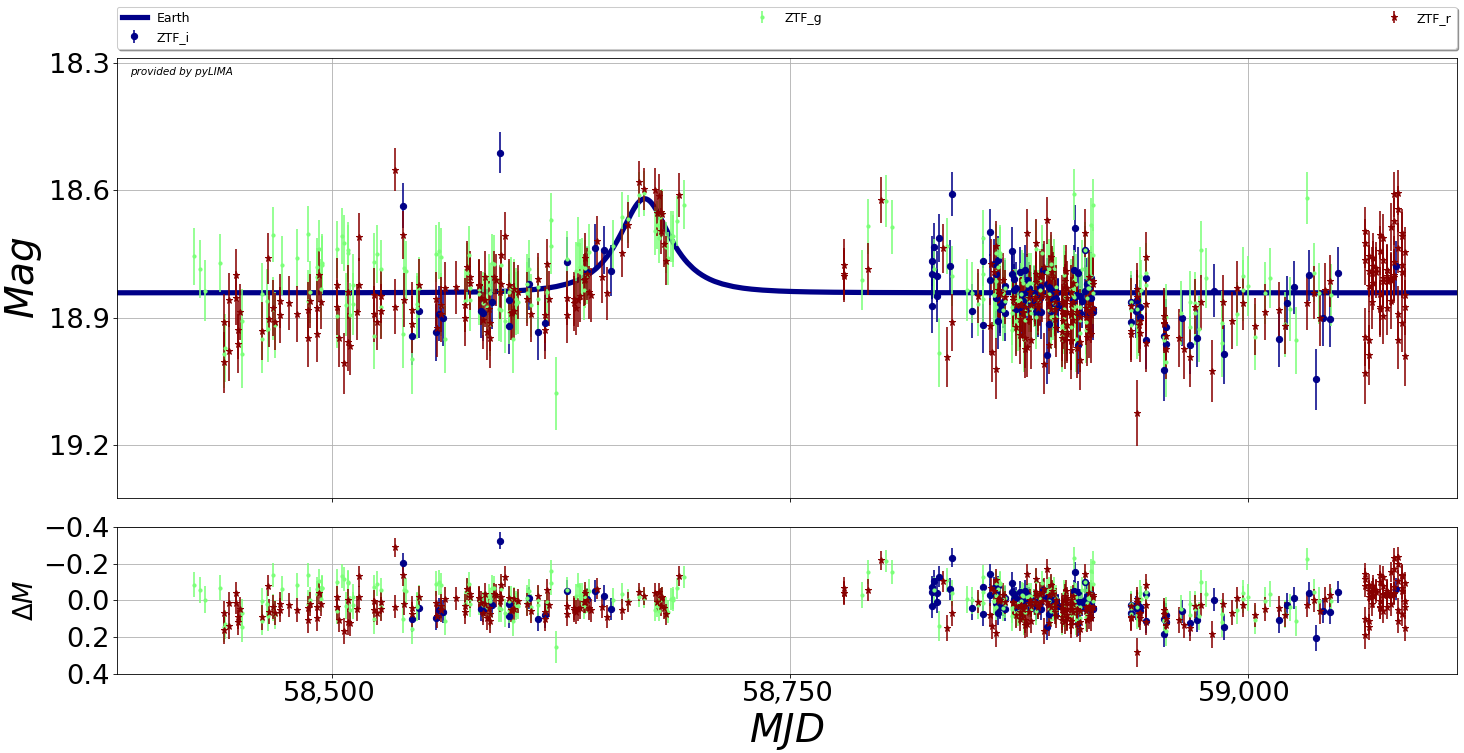}

\caption{(\textbf{Upper panel}): {Light} 
curve of the AGN J1249+3449 fitted with the PSPL model and the model residuals with respect to the best fit, with~the vertical axis showing the magnitude of the source. (\textbf{Lower panel}): Close-up of the light curve and the model residuals, with~the magnitude expressed on the vertical axis. For~both panels, the~horizontal axis represents the time expressed in MJD.
Colors follow the definition given in the caption of Figure~\ref{USBL_fit}. All the magnitudes are intrinsically corrected, with~respect to a common baseline, by~pyLIMA.
\label{PSPL_Fit}}
\end{figure}

\begin{figure}[H]
\includegraphics[scale = 0.275]{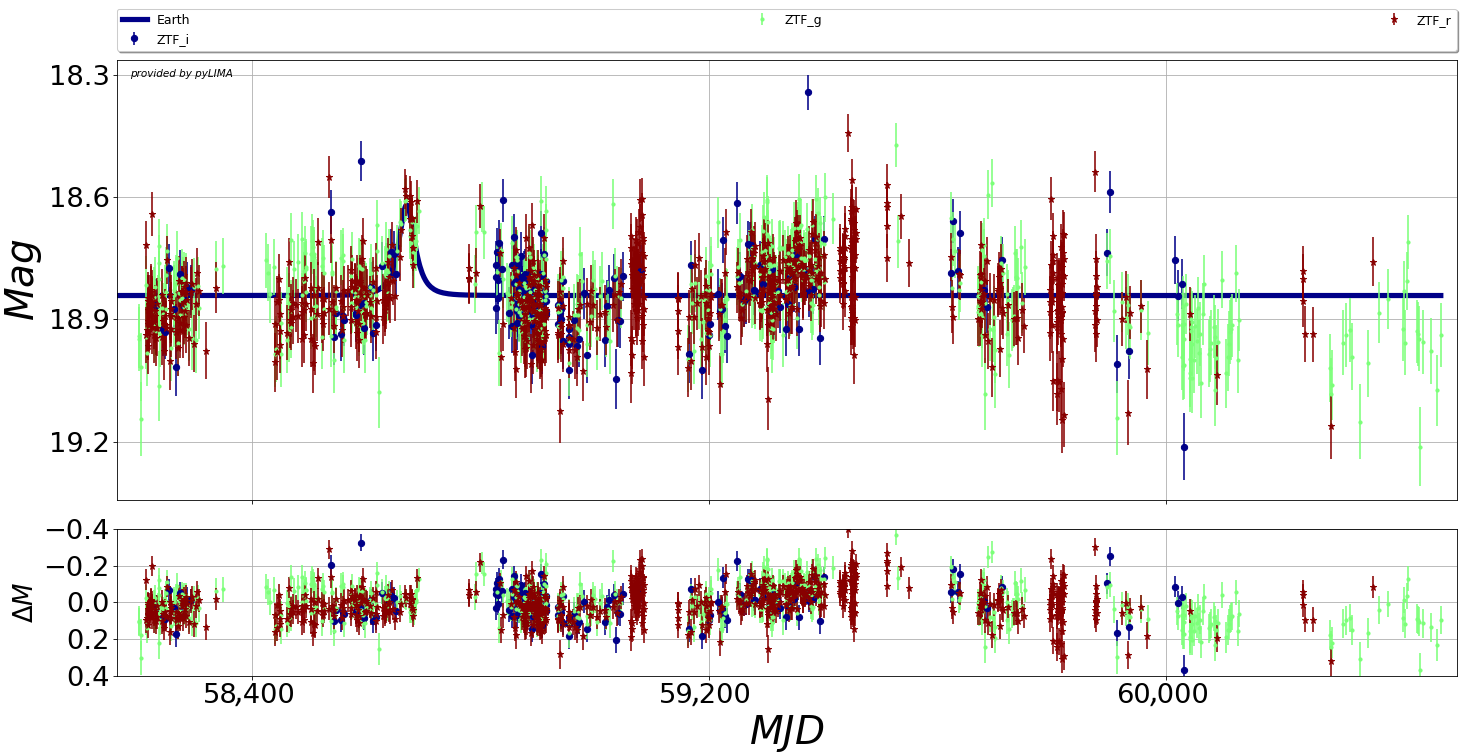}
\\\vspace{-8pt}

{\includegraphics[scale = 0.275]{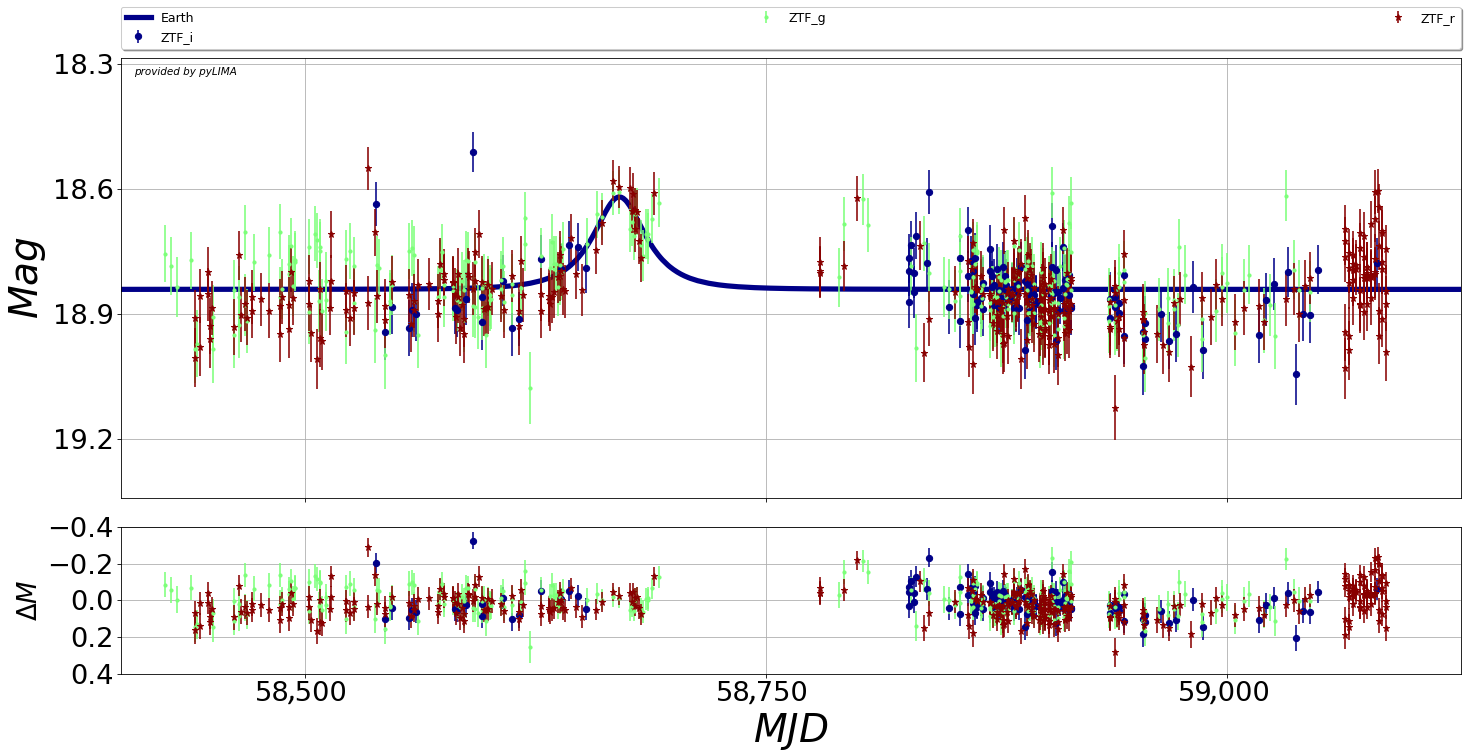}}    
\caption{(\textbf{Upper panel}): {Light} 
curve of the AGN J1249+3449 fitted with the FSPL model and the model residuals with respect to the best fit, with~the vertical axis showing the magnitude of the source. (\textbf{Lower Panel}): Close-up of the light curve and the model residuals, with~the vertical axis representing the magnitude. For~both panels, the~horizontal axis represents the time expressed in MJD. Colors follow the definition given in the caption of Figure~\ref{USBL_fit}. All the magnitudes are intrinsically corrected, with~respect to a common baseline, by~pyLIMA.
\label{FSPL_Fit}}
\end{figure}

Figure~\ref{USBL_fit}, showing the light curve fitted with the USBL model, is characterized by two additional bumps after the main event, which are not predicted in the PSPL and FSPL models (see Figures~\ref{PSPL_Fit} and \ref{FSPL_Fit}).

\unskip

Even though the fit presents a smaller value of the $\chi^2$, the~previous considerations, and~the fact that the value of $a_{\perp}$ does not belong to $1-100\, \text{AU}$, the~typical separation interval of binary systems, allow us to conclude that the USBL model cannot properly account for the quasar light~curve. 

{In} conclusion, the~previous discussion reveals that the PSPL model provides the best description of the light curve of AGN J1249+3449, and~the most likely point-like lens mass turns out to be {$\sim\,$0.1 $M_{\odot}$}.

\section{Conclusions}
The first studies of the flare event towards AGN J1249+3449 were conducted by Graham and colleagues~\cite{Phys.Rev.Lett. 1}, who reported the discovery of a flare-like structure in the quasar's light curve, affirming that this could be the sign of an optical electromagnetic counterpart to a binary black hole merger event. In~the same year, the~research group led by De Paolis reanalyzed the light curve and found that the flare-like signal could be associated with a microlensing event (see~\cite{Mon. Not. R. Astron. Soc. 1}). 

{In} the present paper, the~light curve is examined using significantly more data than in the previous analyses. What emerges is a confirmation of the microlensing hypothesis. 
Indeed, the~light curve is fitted with various microlensing models (the PSPL model, the~FSPL model and the USBL model), showing that the event respects the three minimal requirements for claiming the microlensing nature (see Section~\ref{Light_curve_analysis}). The~most convincing model is the PSPL model, which returns a lens with a mass of {$\sim\,$0.1 $M_{\odot}$}, in~agreement with the value previously found in \citep{Mon. Not. R. Astron. Soc. 1}. 

{With} $0.08\,M_{\odot}$ being the minimum mass for a star, the~object acting as a lens could be either a brown dwarf or a low-mass star. This is another demonstration of the extreme ``sensibility'' of the gravitational microlensing technique to catch effects caused by very light objects. Consequently, gravitational microlensing also permits the detection of these kinds of objects, which, otherwise, are impossible to detect in faraway~galaxies.



\vspace{6pt} 





\authorcontributions{Conceptualization, M.C., F.D.P., A.F. and A.N.; methodology, F.D.P. and A.N.; software, M.C. and A.F.; validation, M.C., A.F. and A.N.; formal analysis, M.C. and A.F.; data curation, M.C. and A.F.; writing---original draft, M.C.; writing---review and editing, F.D.P., A.F. and A.N.; visualization, A.F. and A.N.; supervision, F.D.P. and A.F. All authors have read and agreed to the published version of the manuscript.}

\funding{This research received no external~funding.}

\institutionalreview{Not applicable.}

\dataavailability{The original data presented in the study are openly available in ``ZTF Public Data Releases'' at \url{https://www.ztf.caltech.edu/ztf-public-releases.html} {(accessed on 7 September 2025)}.} 

\acknowledgments{TAsP (Theoretical Astroparticle Physics) and Euclid INFN Scientific Projects are acknowledged. The~two anonymous reviewers are also acknowledged for their~comments.}

\conflictsofinterest{The authors declare no conflicts of~interest.} 

\abbreviations{Abbreviations}{
The following abbreviations are used in this manuscript:
\\

\noindent 
\begin{tabular}{@{}ll}
AGN & Active Galactic Nucleus\\
GW &Gravitational Wave \\
ZTF &Zwicky Transient Facility  \\
BBH & Binary Black Hole\\
PSPL  & Point Source Point Lens \\
  LM&  Levenberg--Marquardt\\
  DE&  Differential Evolution\\
  MCMC& Markov Chain Monte Carlo \\
FSPL  & Finite Source Point Lens \\
  USBL&Uniform Source Binary Lens\\
  AU& Astronomical Unit \\
MJD  & Modified Julian Date \\
 SMBH &Supermassive Black Hole  \\
\end{tabular}
}


\begin{adjustwidth}{-\extralength}{0cm}
\printendnotes[custom] 

\reftitle{References}

\PublishersNote{}
\end{adjustwidth}
\end{document}